\documentclass[a4paper,dvips 2t,superscriptaddress,showkeys,nofootinbib]{revtex4} 
\usepackage{epsfig,graphics,graphicx,amsmath,amssymb}

\usepackage{graphics}
\usepackage{subfigure}
\usepackage{amssymb}
\usepackage{color}
\usepackage{amsmath}
\usepackage{amsthm}
\usepackage{amsopn}
\newcommand\varpm{\mathbin{\vcenter{\hbox{%
				\oalign{\hfil$\scriptstyle+$\hfil\cr
					\noalign{\kern-.3ex}
					$\scriptscriptstyle({-})$\cr}%
}}}}
\newcommand\varmp{\mathbin{\vcenter{\hbox{%
				\oalign{\hfil$\scriptstyle-$\hfil\cr
					\noalign{\kern-.3ex}
					$\scriptscriptstyle({+})$\cr}%
}}}}

\def\uno{\mbox{1 \kern-.59em {\rm l}}}

\def\th{\theta}

\def\be{\begin{equation}}
\def\ee{\end{equation}}
\def\bea{\begin{eqnarray}}
\def\eea{\end{eqnarray}}

\begin{document}
\title{\bf\bf Quantum coherence and entanglement in neutral-current neutrino oscillation in matter}

\author{ M.M. Ettefaghi }
\email{mettefaghi@qom.ac.ir}
\affiliation{Department of Physics, University of Qom, Ghadir Blvd., Qom 371614-6611, I.R. Iran} 
\author{ Z. Askaripour Ravari}
\affiliation{Department of Physics, Islamic Azad University, North Tehran Branch, Tehran, I.R. Iran}

\begin{abstract}
Although neutrino-antineutrino states originating from neutral-current interactions are blind concerning the flavor state, an oscillation pattern is predicted provided that both neutrino and antineutrino are detected. This issue arises from both the coherence and entanglement of the neutrino-antineutrino states. Based on quantum resource theory, we use the $l_1$-norm and concurrence to quantify quantum coherence and entanglement, respectively. Considering the localization properties by the wave packet approach and a matter potential which appears when neutrino and antineutrino propagate in a material medium, we obtain the $l_1$-norm and concurrence. We see that the neutrino and antineutrino remain entangled for larger baseline lengths when they propagate in a material medium. In the case of the coherence property, the $l_1$-norm decreases in comparison to the corresponding one in vacuum. However, its damping occurs for larger distances.

\end{abstract} 

\maketitle

\section{Introduction}\label{1111}
Quantum coherence and entanglement are two important quantum resources that play crucial roles in quantum information and quantum computation science and they are investigated in various branches of condensed matter, atomic physics, and quantum optics. Additionally, several particle physics phenomena, such as meson mixing and neutrino oscillation, could potentially provide suitable opportunities for their explorations \cite{LG2016,c1,b1,b2,epl,e2,e3,e4,epjp,prd,11,12,13}. For instance, violation of the Leggett-Garg inequality as a manifestation of quantum coherence has been investigated using MINOS data \cite{LG2016}.The $l_1$-norm value as a measure of quantum coherence has also been studied using data from the Day Bay, KamLAND, MINOS and T2K neutrino experiments \cite{c1}. For instance, in the case of the Daya Bay experiment, in which detectors are near the source and the measured  transition probability is low, the $l_1$-norm takes a value much lower than its maximum, while in the case of KamLAND, in which the neutrino baseline is about 180 km, this coherence measure reaches the maximal value.
Certainly, the neutrino oscillation process, which is a quantum mechanical phenomenon occurring on the macroscopic scale, is an appropriate medium in which to study the quantum foundation aspects and quantum correlations such as quantum coherence and entanglement.  Using the wave packet approach, one can give a meticulous description of neutrino oscillation \cite{43,44,45,46}. Specifically, the production, propagation
and detection of a neutrino should be considered as
localized processes, and this localization is very well
fulfilled using the wave packet approach. To be more precise, we must emphasize that the observation of neutrino oscillation depends on the coherence of neutrinos during production, propagation, and detection processes.
The production and detection coherence conditions are satisfied provided that the intrinsic quantum
mechanical energy uncertainties during these processes are large compared to the energy difference $\Delta E_{ij}$ of different neutrino mass eigenstates:
\be
\Delta E_{ij}\sim\frac{\Delta m^2_{ij}}{2E}\ll\sigma_E,
\ee
where $\sigma_E=\mbox{min}\{\sigma^{\text{prod}}_E,\sigma^{\text{det}}_E\}$. This condition implies that during the production and detection processes, one cannot discriminate the neutrino mass eigenstates.
Conservation of coherence during propagation means that wave packets describing the mass eigenstates overlap from the production to the detection regions. The wave packets describing different neutrino mass eigenstates propagate with different group velocities.
After propagating $L$, the separation of different mass wave packets is $\frac{\Delta m_{ij}^2}{2E^2}L$.  Consequently, coherent propagation is guaranteed provided that
\be
\frac{\Delta m^2_{ij}}{2E^2}L\ll\sigma_{x\nu}\simeq\frac{v_g}{\sigma_E},\label{eq2}
\ee
where $v_g$ is the average group velocity of wave packets of different
neutrino mass eigenstates, and $\sigma_{x\nu}$ is their common effective spatial width.
In other words, similar to the double-slit experiment, if one could determine which mass eigenstate is created or detected, the neutrino oscillation pattern would disappear. In particular, the conservation of energy and momentum implies that exact determination of the energy-momentum of charged leptons leads to determination of the mass eigenstate of the corresponding neutrino (in fact, exact momentum conservation causes the neutrino state to be kinematically entangled with the corresponding charged lepton state), and the neutrino oscillation ceases \cite{found,glashow}.

Furthermore, the neutrino oscillation process is observed provided that neutrinos have a specific initial flavor state. The kinematic analysis shows that a neutrino state created through charged current (CC) interactions exhibits this situation. For instance, a muon neutrino is created by pion decay while muon decay only gives an electron neutrino. In contrast, the neutral current (NC) processes or the $Z_0$ decay are blind with respect to the neutrino flavor states. In other words, every flavor eigenstate and every mass eigenstate is created with equal probability.
Therefore, one can write, in general, the state of the created neutrino and antineutrino as follows:
\be
|\nu_Z\rangle=\frac{1}{\sqrt{3}}\sum_{\alpha=e,\mu,\tau}|\nu_\alpha\rangle|\bar{\nu}_\alpha\rangle=\dfrac{1}{\sqrt{3}}\sum _{i=1}^3\left|  \nu _i\right\rangle \left|  \bar\nu _i\right\rangle.\label{eq3}
\ee
Here, the second equality is satisfied because the transformation matrix between mass and flavor eigenstates is unitary.
Thus, the usual neutrino oscillation in which either neutrino or antineutrino is detected cannot be observed for NC neutrinos. However, Eq. (\ref{eq3}) shows that there is another property that is noticeable: the neutrino and antineutrino originating from the NC interaction are maximally entangled due to the conservation of energy-momentum and lepton number in electroweak interactions.
 Hence, if both neutrino and antineutrino are detected, it is possible that oscillation patterns will be observed between detectors \cite{sza,4,scrip}. However, if only either neutrino or antineutrino is detected, we will have an entirely classical ensemble and will be unable to observe any oscillation pattern.
 Indeed, this phenomenon can be interpreted as a CPT transformation of half of the process maps of the propagation of neutrino-antineutrino pairs onto the traditional neutrino oscillation experiment. Of course, its requirement is that the initial state is an entangled state according to Eq. (\ref{eq3}). Otherwise, if the initial state is separable as follows:
 \begin{equation}
 	\rho=\frac{1}{3}\sum_{\gamma}|\nu_\gamma\rangle|\bar \nu_\gamma\rangle\langle\nu_\gamma|\langle\bar \nu_\gamma|,
 \end{equation}
using the CPT transformation on the antineutrino process, one can show that the probability of detecting a neutrino with flavor state $|\nu_\alpha\rangle$ at a detector and an antineutrino with flavor state $|\nu_\beta\rangle$ at another detector is
\begin{equation}
	P_{\alpha\beta}=\frac{1}{3}\sum_{\gamma}P_{\gamma\alpha}P_{\beta\gamma},
\end{equation}
which is not the traditional neutrino oscillation probability.
 Although NC neutrino oscillation is beyond the reach of any experiment, it can be implemented analogically by the double-double-slit experiment performed in Ref. \cite{dds}. In this experiment, the path-entangled photons are passed through opposite screens with a double slit. Due to entanglement, each photon can reveal the which-slit-path information of the other photon. Therefore, the two-photon interference pattern appears if detection locations of photons are correlated without revealing the which-slit-path information. However, it has also been shown experimentally and theoretically that two-photon quantum interference disappears when the which-slit-path of one photon is detected in the double-double-slit. 
  
 When neutrinos (antineutrinos) propagate in a material medium, they might have a forward coherent scattering off electrons and nucleons of the matter via both weak CC and weak NC interactions \cite{wolf, mikh}. All three neutrino (antineutrino) flavor states can interact with the matter through NC, but the electron neutrino (antineutrino) has an additional interaction (i.e. interaction through the CC) which causes it to feel an additional potential. In the calculations, we only consider the effective potential due to the CC interaction because the potentials coming from the NC interactions induce common phases for all three neutrino (antineutrino) flavor states and do not alter neutrino oscillating behaviors. Therefore, the effective potential of matter which must be considered is $V_{CC}=\sqrt{2}G_FN_e$ ($\bar V_{CC}=-\sqrt{2}G_FN_e$), where $G_F$ and $N_e$ are the Fermi coupling constant and the density of electrons in the medium, respectively. If the matter density of the medium is nonuniform, the potential depends on the coordinate; otherwise, it is constant. 
  
  According to Eq. (\ref{eq3}), the neutrino and antineutrino produced by NC interactions are entangled with respect to either their flavor or mass modes. Indeed, this state has a maximal entanglement similar to the Bell states\footnote{Bell states are the four states that can be created when two qubits are maximally entangled. The four states are represented as
 $
  	|\Phi^\pm\rangle=\frac{|00\rangle\pm|11\rangle}{\sqrt{2}}$ and $|\Psi^\pm\rangle=\frac{|10\rangle\pm|10\rangle}{\sqrt{2}}.
$}. Moreover, they are entangled due to energy-momentum conservation. In fact, the predicted oscillation pattern between two detectors is based on these features.  Also, as a result of entanglement, the coherence condition obtained by considering localization properties via the wave packet approach is stronger than that obtained by imposing the non-separation mass wave packet constraint analogous to Eq. (\ref{eq2}) \cite{4,scrip}. In fact, the coherent propagation of both neutrino and antineutrino is not sufficient because the oscillation pattern ceases if the distance between the detectors is larger than the coherence length. In this paper, we consider the matter refraction to reanalyze this problem in the context of two-flavor neutrino oscillations. In this case, the NC neutrino oscillation cannot be concluded by a CPT transformation of half the process. However, since the mixing in the source is independent of matter potential, the initial state can be written in mass eigenstate according to Eq. (\ref{eq3}). 
Therefore, as we will show, one can obtain the standard oscillation pattern in a material medium for the NC neutrinos.
  Moreover, using quantum resource theory, we study the quantum entanglement and quantum coherence under these conditions. For this purpose, concurrence and $l_1$-norm are defined as a measure for entanglement and quantum coherence, respectively. Coherence is the quantum resource of neutrino oscillation, and its origin during production and detection processes, as well as its conservation during propagation, has been discussed extensively and accurately in the literature; for instance, see \cite{43,44,45,46,found,glashow,a2012,guintidensity}. Meanwhile, using quantum resource theory, one can quantify quantum coherence, entanglement, or another quantum resource by introducing some measure quantities \cite{Vedral,Horodecki,hill,wootters,bau,str}. In the case of usual neutrino oscillation, the refraction of the neutrino in a dense medium leads to a decrease in the quantum coherence in standard neutrino oscillation \cite{epjp}. However, as we will see, the value of entanglement of neutrino-antineutrino pairs shows different behavior in terms of matter potential.

In the next section, treating the neutrino and antineutrino as a wave packet, we obtain the probability of the NC neutrino oscillation in a material medium with constant density for the two-flavor schema.  In section \ref{333}, we explore the variation in entanglement and quantum coherence for this issue in the quantum resource theory framework. For this purpose, we use concurrence and $l_1$-norm measures for entanglement and coherence, respectively. We summarize our results in the last section.

\section{Probability of NC neutrino oscillation in matter}\label{2222}
According to the quantum wave packet approach, we should describe the neutrino and antineutrino by a localized wave function at the space-time coordinates $(x,t)$ and $(\bar x,\bar t)$, respectively. Moreover, it is assumed that the neutrino and antineutrino detectors are located at distances $L$ and $\bar L$ from the source, respectively.  The neutrino and antineutrino coming from the $Z_0$ boson must be described by a bipartite entangled state.
Therefore, this state can be written as follows:
\be\label{so}
\left|  \nu _Z,{\bf x},t,{\bf\bar x},\bar t\right\rangle = \dfrac{1}{\sqrt{2}}\sum_i \Psi_{iS}^m({\bf x},t,{\bf\bar x},\bar t)\left|  \nu _i\right\rangle \left|  \bar\nu _i\right\rangle,
\ee
where
\bea\nonumber\label{psi}
\Psi_{iS}^m({\bf x},t,{\bf\bar x},\bar t)&=&\textit{N}\int\dfrac{d^3p}{(\sqrt{2\pi})^3}\int\dfrac{d^3\bar p}{(\sqrt{2\pi})^3} f_S ({\bf p},{\bf p}^m_i) \bar f_S({\bf\bar p},{\bf\bar p}^m_i)\delta^3({\bf p}-{\bf \bar p})\\
&\times&\exp[-iE_i^m({\bf p})(t-t_p)+i{\bf p}({\bf x}-{\bf x_p})-i\bar E^m_i({\bf\bar p})(\bar t-t_p)+i{\bf\bar p}\cdot({\bf\bar x}-{\bf x_p})].
\eea
The superscript $m$ refers to the propagation in a material medium, and the subscript $S$ indicates that this state is related to the source. Here, $f_S ({\bf p},{\bf p}^m_i)$ and $\bar f_S({\bf\bar p},{\bf\bar p}^m_i)$ are the momentum distribution functions of the neutrino and antineutrino with mean momenta ${\bf p}^m_i$ and ${\bf\bar p}^m_i$. 
Also, $E^m_i({\bf p})$ and $\bar E^m_i({\bf \bar p})$ are the energies of neutrino and antineutrino with mass $m_i$, respectively, and $t_p$ and ${\bf x}_p$ denote the production coordinates.
 The Dirac delta factor is added in Eq. (\ref{psi}) to ensure that the momentum is conserved (here, we consider those $Z_0$ bosons as being almost at rest in the frame where the matter effect is constant). Indeed, the time evolution of neutrino states in a matter medium is governed by the following effective Hamiltonian \cite{1}:
\be
\delta H^m=H_0 + V_{CC}=\frac{1}{2E}[{U}{\text{diag}}(0,\Delta m_{21}^2){U}^\dagger+{\text{diag}(2EV(x),0)}].\label{H}
\ee
In the case of antineutrino, we denote the corresponding Hamiltonian by $\delta{\bar H}^m$, and it is obtained from Eq. (\ref{H}) by replacing $V(x)$ with $-V(x)$. We consider the density of matter to be constant. The unitary mixing matrix $U$ for a two-generation platform is parameterized as follows:
\be\label{mix}
U=\begin{bmatrix}
	\cos \theta & \sin\theta \\
	-\sin\theta&\cos \theta \\
\end{bmatrix}.
\ee
 Therefore, one can obtain the following eigenvalues for the Hamiltonian given by Eq. (\ref{H}):
\be
\delta E_1^m=\dfrac{\Delta m_{21}^2 +2EV-\sqrt{4E^2V^2+\Delta m_{21}^4-4EV\Delta m_{21}^2\cos2\theta}}{4E},\label{e1}
\ee
and
\be
\delta E_2^m=\dfrac{\Delta m_{21}^2 +2EV+\sqrt{4E^2V^2+\Delta m_{21}^4-4EV\Delta m_{21}^2\cos2\theta}}{4E},\label{e2}
\ee
and the corresponding mixing angle is given by
\be\label{t12}
\theta^m=\dfrac{1}{2}\arctan\left(\dfrac{\Delta m_{21}^2\sin 2\theta}{\Delta m_{21}^2\cos2\theta-2EV}\right).
\ee
In the case of the antineutrino, we denote the eigenvalues of $\delta{\bar H}^m$ and corresponding mixing angle by ${\delta\bar E}^m_{1,2}$ and ${\bar \theta}^m$, respectively. These parameters can be represented by the corresponding expressions given in Eqs. (\ref{e1}), (\ref{e2}) and (\ref{t12}) with the difference that $V$ must be replaced by $-V$. It is clear from Eq. (\ref{t12}) that there is a resonance in mixing of neutrinos provided that $V\simeq\Delta m_{21}^2\cos2\theta/2E$. However, no resonance exists for antineutrino mixing.

For the momentum distribution function of the neutrino and antineutrino in Eq. (\ref{psi}), we use the Gaussian momentum wave function as follows: 
\be\label{gauss}
f({\bf p,p}_i)=\left(\dfrac{2\pi}{\sigma_p^2}\right)^{\frac{3}{4}}\exp\left[-\dfrac{({\bf p-p}_i)^2}{4\sigma_p^2}\right].
\ee
 Hereafter, the momentum uncertainties of the neutrino and antineutrino are denoted by $\sigma_p$ and ${\bar \sigma}_p$, respectively.
 It is appropriate to assume that $\sigma_p$ and ${\bar \sigma}_p$ are much smaller than the corresponding mean momenta. Therefore, one can expand the energy up to the first order of the departure from the mean momentum as follows:
\be\label{energym}
E_i^m({\bf p})\simeq E_i^m+{\bf v}_i^m({\bf p}-{\bf p}_i^m),
\ee
where
\be
E_i^m\equiv E_i^m({\bf p}_i^m)=\sqrt{m_i^2+{ {\bf p}_i^m}^2},
\ee
and  ${\bf v}^m_i$ is the group velocity of the $i$th mass eigenstate and is given by ${\bf v}^m_i=dE^m/d{\bf p}|_{{\bf p}={\bf p}^m_i}$. Likewise, we can write a similar expression for antineutrinos.
Consequently, one can write the position wave packet given in Eq. (\ref{psi}) as follows:
\bea\nonumber
\Psi_{iS}^m({\bf x},t,\bar {\bf x},\bar t)&\propto&\exp\left[-i(E_i^mt+\bar E_i^m \bar t)-\dfrac{(\sigma_{x\nu}^2\bar\sigma_{x\bar \nu}^2)}{(\sigma_{x\nu}^2+\bar\sigma_{x\bar\nu}^2)}({\bf p}_i^m-\bar {\bf p}_i^m)^2-\dfrac{({\bf x}+\bar {\bf x}-{\bf v}_i^mt-\bar {\bf v}_i^m\bar t)^2}{4(\sigma_{x\nu}^2+\bar\sigma_{x\bar\nu}^2)}\right.\\
&+&\left.\dfrac{i}{(\sigma^2_{x\nu}+\bar\sigma^2_{x\bar\nu})}({\bf x}+\bar {\bf x})(\bar\sigma^2_{x\bar\nu}\bar {\bf p}_i^m+\sigma^2_{x\nu} {\bf p}_i^m)+\dfrac{i}{(\sigma^2_{x\nu}+\bar\sigma^2_{x\bar\nu})}({\bf p}_i^m-\bar {\bf p}_i^m)(\bar\sigma^2_{x\bar\nu} {\bf v}_i^mt-\sigma^2_{x\nu}\bar {\bf v}_i^m\bar t)\right].\label{psi1}
\eea
where $\sigma^2_{x\nu}$ and $\bar \sigma^2_{x\bar \nu}$ are the position uncertainties of the neutrino and antineutrino in the  source, and they are given by $\sigma_{x}=1/2\sigma_p$ and $\bar \sigma_{x}=1/2\bar \sigma_p$, respectively.

On the other hand, since the detection processes are essentially time-independent, detected states have no time
dependence. Therefore, the wave function of the detected neutrino and antineutrino states is described by
\be\label{dete}
\left|  \nu _\alpha,{\bf x-L},\bar\nu_\beta,{\bf\bar x-\bar L}\right\rangle = \sum_i\sum_j U_{\alpha i}^{m*}\bar U_{\beta j}^m\Psi_{iD}^m({\bf x-L})\bar\Psi^m_{j\bar D}({\bf\bar x-\bar L})\left|  \nu _i\right\rangle \left|  \bar\nu _j\right\rangle,
\ee
where $\Psi_{iD}^m({\bf x-L})$ and $\bar\Psi_{j\bar D}^m({\bf\bar x-\bar L})$ are the wave functions of the detected neutrino and antineutrino at positions ${\bf L}$ and $\bar {\bf L}$, respectively. Moreover, $U_{\alpha i}^{m}$ and $\bar U_{\beta i}^m$ in Eq. (\ref{dete}) are mixing matrices of neutrinos and antineutrinos in a material medium. In fact, $U^{m}$ ($\bar U^m$) has a form similar to Eq.(\ref{mix}) with the difference that $\theta$ is replaced by $\theta^m$ ($\bar \theta^m$)  given by Eq. (\ref{t12}). Again, we assume a localized Gaussian
wave function in momentum space similar to (\ref{gauss}) for both the neutrino and antineutrino detected in the corresponding detectors.
Under this scenario, the detected state Eq.(\ref{dete}) can be written as follows
\be\label{de}
\left|  \nu _\alpha,{\bf x-L},\bar\nu_\beta,{\bf\bar x-\bar L}\right\rangle = \sum_{i}U_{\alpha i}^{m*}\bar U_{\beta j}^m\exp\left[i{\bf p}'^m_i({\bf x-L})+i{\bf\bar p}'^m_i({\bf\bar x-\bar L})-\dfrac{({\bf x-L})^2}{4\sigma^2_{xD}}-\dfrac{({\bf\bar x-\bar L})^2}{4 \bar\sigma^2_{x\bar D}}\right]\left|  \nu _i\right\rangle \left|  \bar\nu _j\right\rangle,
\ee
where $\sigma^2_{xD}$ and $\bar \sigma^2_{x\bar D}$ are the uncertainties of the detected neutrino and antineutrino processes. In general, the mean momenta of the produced particles, ${\bf p}_i$ and ${\bf\bar p}_i$, are different from one of the detected particles, ${\bf p}'_i$ and ${\bf\bar p}'_i$. However, we assume that they coincide.
Then, the transition amplitude of detecting $\nu_\alpha$ and  $\bar\nu_\beta $ in the corresponding detectors becomes:
\be
A^m_{\alpha\beta}=\int d^3x\int d^3\bar x\langle \nu_\alpha,{\bf x-L}, \bar\nu_\beta,{\bf\bar x-\bar L} \mid \nu _Z,{\bf x},t,{\bf\bar x},\bar t\rangle.
\ee
Substituting Eq.(\ref{so}) with $\Psi_{iS}^m$ given in Eqs. (\ref{psi1}) and (\ref{de}) into the above equation and calculating the integrals over the position coordinates, we obtain
\bea\nonumber
A^m_{\alpha\beta}\propto\dfrac{1}{\sqrt{2}}\sum_{i}U_{\alpha i}^{m*}\bar U_{\beta i}^m\!\!\!&\text{exp}&\!\!\!\left[-iE_i^mt-i\bar E_i^m\bar t-\dfrac{({{\bf L}+\bar{\bf L}}-{\bf v}_i^m t-{\bf\bar v}_i^m\bar t)^2}{8\sigma_x^2}-\dfrac{\sigma^2_x({\bf p}_i^m-\bar {\bf p}_i^m)^2}{2}\right.\\
&+&\!\!\!\!\left.\dfrac{i({\bf L}+\bar {\bf L})({\bf p}_i^m+\bar {\bf p}_i^m)}{2}+\dfrac{i({\bf p}_i^m-\bar {\bf p}_i^m)({\bf v}_i^mt-\bar {\bf v}_i^m\bar t)}{2}\right],
\eea
 where we assumed that the position uncertainties of neutrinos and antineutrinos which  are defined  as
\be
\sigma_{x}^2\equiv\sigma_{x\nu}^2+\sigma_{xD}^2,\,\,\,\,\,\,\bar\sigma_{x}^2\equiv\bar\sigma_{x\bar\nu}^2+\bar\sigma_{x\bar D}^2,
\ee
are approximately equal $\sigma_{x}^2\simeq\bar\sigma_{x}^2$.

We want to obtain the probability of detecting a neutrino with flavor state $|\nu_\alpha\rangle$ at one detector and the corresponding antineutrino with flavor state $|\nu_\beta\rangle$ at another. 
In general, the distance of the two detectors from the production location may be different. In this case, the arrival times of the neutrino and antineutrino at their corresponding detectors are not equal. It is suitable that instead  of $t$ and $\bar t$, we select the following time variables:
	\begin{equation}
	T=t+\bar t, \hspace{1cm}\text{and} \hspace{1cm} \tau=|t-\bar t|.
	\end{equation}
 Meanwhile, the emission time and
the corresponding arrival times of the neutrino and antineutrino are not measured. Hence, we integrate over $T$, which is the arrival time of the neutrino or antineutrino from its own detector to another detector.
Therefore, the probability of the oscillation pattern between detectors can be obtained as follows: 
\bea\nonumber
P^m_{\alpha\beta}&\propto&\dfrac{1}{2}\sum_{i,j}U_{\alpha i}^{m*}U_{\alpha j}^{m}\bar U_{\beta j}^{m*}\bar U_{\beta i}^m\,\,\,\text{exp}\Bigg\{-\dfrac{({\bf L}+\bar{\bf  L})^2}{8\sigma_{ x}^2}\dfrac{(\Delta {\bf v}_{ij}^m+\Delta\bar {\bf v}_{ij}^m)^2}{({\bf v}_i^m+\bar {\bf v}_i^m)^2+({\bf v}_j^m+\bar {\bf v}_j^m)^2}\\\nonumber
&-&\dfrac{2\sigma_x^2(\Delta E_{ij}^m+\Delta\bar E_{ij}^m)^2}	
{({\bf v}_i^m+\bar {\bf v}_i^m)^2+({\bf v}_j^m+\bar {\bf v}_j^m)^2}-\dfrac{\sigma_{x}^2}{2}\bigg[\Big(\bar {\bf p}_i^m-{\bf p}_i^m\Big)^2+\Big(\bar {\bf p}_j^m-{\bf p}_j^m\Big)^2\bigg]\\\nonumber
&-&i\dfrac{({\bf L}+\bar {\bf L})(\Delta E_{ij}^m+\Delta\bar E_{ij}^m)({\bf v}_i^m+{\bf v}_j^m+\bar {\bf v}_i^m+\bar {\bf v}_j^m)}{({\bf v}_i^m+\bar {\bf v}_i^m)^2+({\bf v}_j^m+\bar {\bf v}_j^m)^2}+i\dfrac{({\bf L}+\bar {\bf L})(\Delta {\bf p}_{ij}^m+\Delta\bar {\bf p}_{ij}^m)}{2}\\\nonumber
&+&2\sigma_x^2\dfrac{(\Delta E_{ij}^m+\Delta\bar E_{ij}^m)\Big[(\bar {\bf v}_i^m- {\bf v}_i^m)(\bar {\bf p}_i^m-{\bf p}_i^m)-(\bar {\bf v}_j^m- {\bf v}_j^m)(\bar {\bf p}_j^m-{\bf p}_j^m)\Big]}{({\bf v}_i^m+\bar {\bf v}_i^m)^2+({\bf v}_j^m+\bar {\bf v}_j^m)^2}\\\nonumber
&-& \dfrac{\sigma_x^2}{2}\dfrac{\Big[({\bf v}_i^m-\bar {\bf v}_i^m)(\bar {\bf p}_i^m-{\bf p}_i^m)-({\bf v}_j^m-\bar {\bf v}_j^m)(\bar {\bf p}_j^m-{\bf p}_j^m)\Big]^2}{({\bf v}_i^m+\bar {\bf v}_i^m)^2+({\bf v}_j^m+\bar {\bf v}_j^m)^2} \\\nonumber
&-& i\dfrac{({\bf L}+\bar {\bf L})}{2}\dfrac{({\bf v}_i^m+{\bf v}_j^m+\bar {\bf v}_i^m+\bar {\bf v}_j^m)\Big[(\bar {\bf p}_i^m-{\bf p}_i^m)({\bf v}_i^m-\bar {\bf v}_i^m)-(\bar {\bf p}_j^m-{\bf p}_j^m)({\bf v}_j^m-\bar {\bf v}_j^m)\Big]}{({\bf v}_i^m+\bar {\bf v}_i^m)^2+({\bf v}_j^m+\bar{\bf v}_j^m)^2}\\\nonumber
&-&\dfrac{(\bar{\bf v}_j^m{\bf v}_i^m-\bar{\bf v}_i^m{\bf v}_j^m)^2\tau^2}{8\sigma_x^2(({\bf v}_i^m+\bar {\bf v}_i^m)^2+({\bf v}_j^m+\bar{\bf v}_j^m)^2)}-i\dfrac{\Delta E_{ij}^m\tau}{({\bf v}_i^m+\bar {\bf v}_i^m)^2+({\bf v}_j^m+\bar{\bf v}_j^m)^2}(\bar{\bf v}_i^{m2}+\bar{\bf v}_j^{m2}+\bar{\bf v}_i^m{\bf v}_i^m+\bar{\bf v}_j^m{\bf v}_j^m)\\\nonumber
&+&i\dfrac{\Delta \bar E_{ij}^m\tau}{({\bf v}_i^m+\bar {\bf v}_i^m)^2+({\bf v}_j^m+\bar{\bf v}_j^m)^2}({\bf v}_i^{m2}+{\bf v}_j^{m2}+\bar{\bf v}_i^m{\bf v}_i^m+\bar{\bf v}_j^m{\bf v}_j^m)+\dfrac{({\bf L}+\bar{\bf L})(\bar{\bf v}_i^m{\bf v}_j^m-\bar{\bf v}_j^m{\bf v}_i^m)\tau}{4\sigma_x^2(({\bf v}_i^m+\bar {\bf v}_i^m)^2+({\bf v}_j^m+\bar{\bf v}_j^m)^2)}(\Delta{\bf v}_{ij}^m+\Delta \bar{\bf v}_{ij}^m)\\\nonumber
&+&i\dfrac{\tau}{2(({\bf v}_i^m+\bar {\bf v}_i^m)^2+({\bf v}_j^m+\bar{\bf v}_j^m)^2)}\bigg[\bigg(2\bar{\bf v}_i^m{\bf v}_i^m(\bar{\bf v}_i^m+{\bf v}_i^m)+(\bar{\bf v}_j^m+{\bf v}_j^m)(\bar{\bf v}_j^m{\bf v}_i^m+\bar{\bf v}_i^m{\bf v}_j^m)\bigg)({\bf p}_i^m-\bar{\bf p}_i^m)\\
&-&\bigg(2\bar{\bf v}_j^m{\bf v}_j^m(\bar{\bf v}_j^m+{\bf v}_j^m)+(\bar{\bf v}_i^m+{\bf v}_i^m)(\bar{\bf v}_j^m{\bf v}_i^m+\bar{\bf v}_i^m{\bf v}_j^m)\bigg)({\bf p}_j^m-\bar{\bf p}_j^m)\bigg]
\Bigg\},\label{p1}
\eea
in which $\Delta E_{ij}^m$ ($\Delta\bar E_{ij}^m$) and $\Delta {\bf v}_{ij}^m$ ($\Delta\bar {\bf v}_{ij}^m$) are the energy and group velocity differences in the material medium between the two mass eigenstates of the neutrino (antineutrino), respectively. By using relativistic approximations, one can write the mean energies as follows: 
\be
E_i^m\approx E+\xi\delta{E}_i^m,\label{em}
\ee
and
\be
\bar E_i^m\approx E+\xi\delta\bar{E}_i^m,
\ee
where  $E$ is the neutrino and antineutrino energy in the limit of zero mass. Also, $\xi$ denotes a dimensionless quantity whose value can be estimated from energy-momentum conservation in the production process \cite{2000qm,giu02}. In the case of $Z_0$ boson decay in the rest frame, we have $\xi\approx 0$. Similarly, the corresponding momenta can be given by
\be
{\bf p}_i^m\approx E+(\xi-1)\delta E_i^m,
\ee
and
\be
\bar {\bf p}_i^m\approx E+(\xi-1)\delta\bar E_i^m.
\ee
Furthermore, in this approximation, the group velocity of a mass eigenstate is also written as 
\be\label{v}
{\bf v}_i^m\approx\dfrac{dE_i^m}{dE}=1+\dfrac{d\delta{E}_i^m}{dE},
\ee
for neutrinos and 
\be
\bar {\bf v}_i^m\approx\dfrac{d\bar E_i^m}{dE}=1+\dfrac{d\delta{\bar E}_i^m}{dE},\label{vmbar}
\ee
for antineutrinos. Therefore, Eq. (\ref{p1}) can be simplified as follows:
\bea\nonumber\label{pab}
P^m_{\alpha\beta}(\mathbf{L},\mathbf{\bar L})\propto \dfrac{1}{2}\sum_{i,j}U^{m*}_{\alpha i}{\bar U}^m_{\beta i}U^m_{\alpha j}{\bar U}^{m*}_{\beta j}&\!\!\!\exp\!\!\!&\Bigg\{-2\pi i\dfrac{({\bf L}+\bar {\bf L})}{{{\bf L}^m}_{ij}^{\text{osc}}}-\dfrac{1}{2}\Big(\dfrac{{\bf L}+\bar {\bf L}}{{{\bf L}^m}_{ij}^{\text{coh}}}\Big)^2-\Big(\dfrac{2\pi \xi \sigma_x}{{{\bf L}^m}_{ij}^{\text{osc}}}\Big)^2\\\nonumber
&-\!\!\!\!\!&\dfrac{\sigma_x^2(\xi-1)^2}{2}\bigg[(\delta E_i^m-\delta\bar E_i^m)^2+(\delta  E_j^m-\delta\bar E_j^m)^2\bigg]\Bigg\}\\\nonumber
&\times&\exp\Bigg\{\dfrac{\tau}{64}\bigg[-32i(\Delta\delta E_{ij}^m-\Delta\delta\bar E_{ij}^m)\\
&+&\dfrac{1}{\sigma_x^2}(\Delta\bar{\bf v}_{ij}^m-\Delta{\bf v}_{ij}^m)\Big(2({\bf L}+\bar{\bf L})(\Delta\bar{\bf v}_{ij}^m+\Delta{\bf v}_{ij}^m)-(\Delta\bar{\bf v}_{ij}^m-\Delta{\bf v}_{ij}^m)\tau\Big)\bigg]\Bigg\},
\eea
where the oscillation length ${{\bf L}^m}_{ij}^{\text{osc}}$ and the coherence length ${{\bf L}^m}_{ij}^{\text{coh}}$,  for $i\neq j$, are defined by
\be\label{osc}
{{\bf L}^m}_{ij}^{\text{osc}}\equiv \dfrac{4\pi}{\Delta\delta E^m_{ij}+\Delta\delta\bar E^m_{ij}},
\ee
in which $\Delta\delta E^m_{ij}\equiv\delta E^m_{i}-\delta E^m_{j}$ and $\Delta\delta \bar E^m_{ij}\equiv\delta \bar E^m_{i}-\delta \bar E^m_{j}$, and
\be\label{coh}
{{\bf L}^m}_{ij}^{\text{coh}}\equiv \dfrac{4\sqrt{2}\sigma _x}{\mid\Delta {\bf v}^m_{ij}+\Delta\bar {\bf v}^m_{ij}\mid},
\ee
respectively.
According to Eq. (\ref{v}), the differences in the neutrino group velocities appearing in Eq. (\ref{coh}) become
\be
\Delta {\bf v}_{ij}=-\dfrac{\Delta m_{21}^2(\Delta m_{21}^2 -2EV\cos2\theta)}{4E^3\Delta \delta E^m_{ij}}.
\ee
For $\Delta\bar {\bf v}^m_{ij}$, we obtain a similar expression, with the difference that  $\Delta \delta E^m_{ij}$ and $V$ must be replaced by $\Delta \delta \bar E^m_{ij}$ and $-V$. For neutrinos, $\Delta {\bf v}_{ij}$ becomes zero provided that $V=\Delta m_{21}^2/2E\cos2\th$. But in the case of antineutrinos, it does not happen. Therefore, according to Eq. (\ref{coh}), the greatest value of ${{\bf L}^m}_{ij}^{\text{coh}}$ occurs when  $\Delta {\bf v}^m_{ij}=0$. The last exponential factor appearing in Eq. (\ref{pab}) is due to the non-simultaneous neutrino and antineutrino detection processes. This factor depends on the subtraction of the difference between the energies and group velocities of the neutrino and antineutrino mass eigenstates. This difference is zero in vacuum, and in the material medium with usual densities its effect can be completely ignored compared to the oscillation and decoherence factors. Therefore, with accuracy such that this factor can be ignored, the processes of neutrino and antineutrino detection can be considered to be simultaneous.
Furthermore, the first three terms appearing in the first exponential, in Eq. (\ref{pab}), are the same as the results of Refs. \cite{4,scrip} with the difference that ${{\bf L}^m}_{ij}^{\text{osc}}$ and ${{\bf L}^m}_{ij}^{\text{coh}}$ are modified due to the neutrino and antineutrino scattering off matter. 
However, the last term in this exponential is coming absolutely from the difference between the potential of neutrinos and antineutrinos. Similar to the terms coming from the non-simultaneity of neutrino and antineutrino detection processes, the effect of this term can also be ignored. 

Since the lifetime of $Z_0$ is so small (about $\tau_{Z_0}\simeq 3\times 10^{-25}s$),  the interval between two nearest collisions becomes smaller than the mean distance between particles in the nucleus. Therefore, $\sigma_x$ is determined by the $Z_0$ decay width $\Gamma_{Z_0}$ as follows \cite{45}:
	\begin{equation}
		\sigma_x\simeq\frac{p}{E}\Gamma_{Z_0}^{-1}.
	\end{equation}
The energy of the neutrino and antineutrino coming from the decay of a rest $Z_0$ is about  $E=46 \text{GeV}$. 
If we take  $\Delta m^2_{21}=7.53\times10^{-5}eV^2$, and $\theta=33.46^\circ$  \cite{2020e}, $\sigma_x$ of approximately $10^{-16} m$ is obtained. With these values for relevant parameters, using (\ref{coh}) and (\ref{osc}), coherence and oscillation length of about $10^{10} m$ and $10^{15} m$ are obtained, respectively. Therefore, under this condition, the oscillation pattern ceases due to the separation of mass eigenstate wave packets. This situation is not improved by choosing the oscillation parameters between the second and third generations, and the oscillation length will still be several orders of magnitude larger than the coherence length. In the following, however, we want to investigate quantum coherence and entanglement as a resource of the oscillation pattern in this problem. Therefore, we suppose $\sigma_x$ to be about the atomic distance ($5\times 10^{-10}\text{m}$) so that the oscillating behavior does not cease.
\begin{figure}[ht]
	\centering
	\subfigure[]{\includegraphics[scale=0.5]{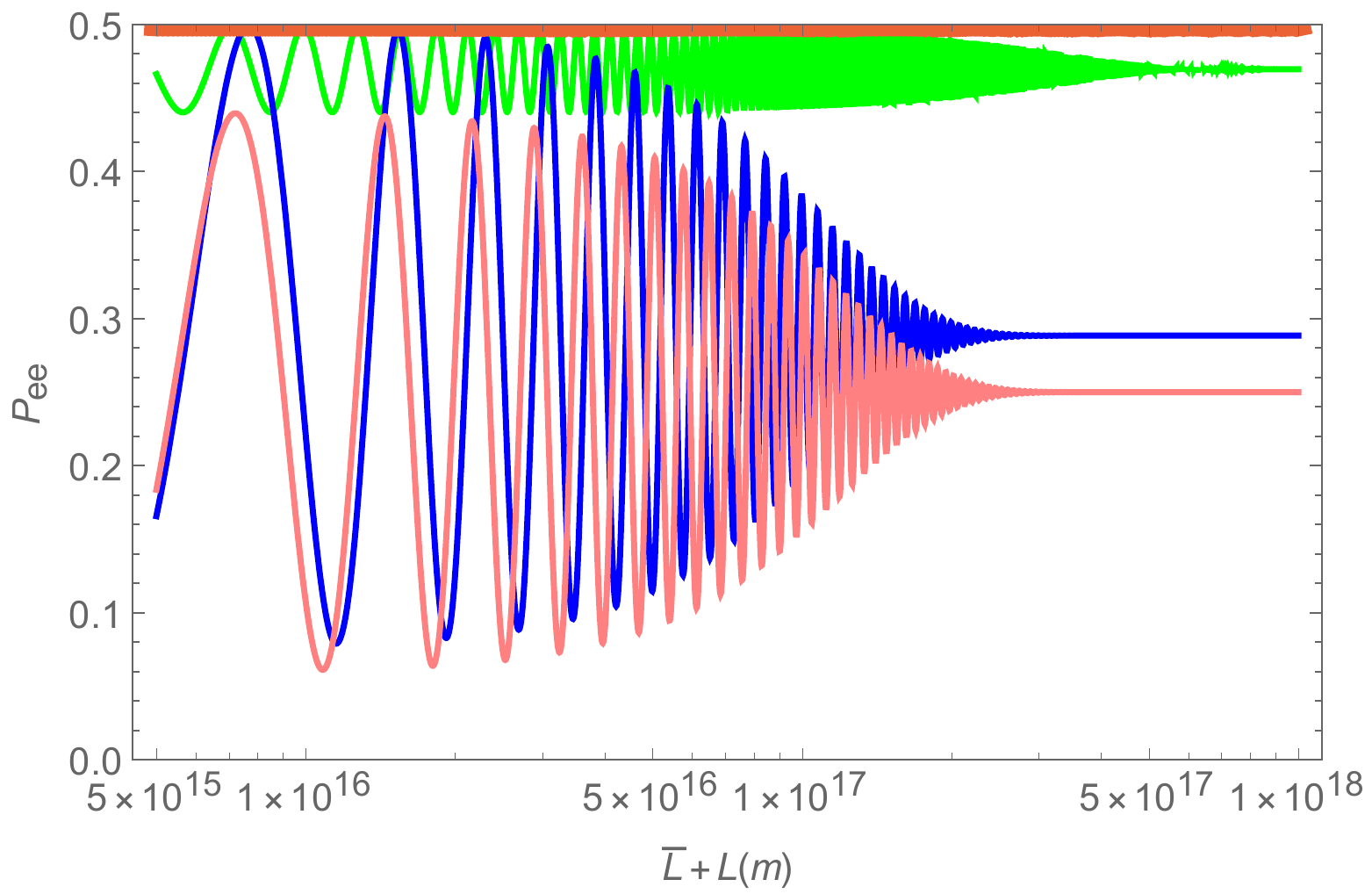}}\subfigure[]{\includegraphics[scale=0.5]{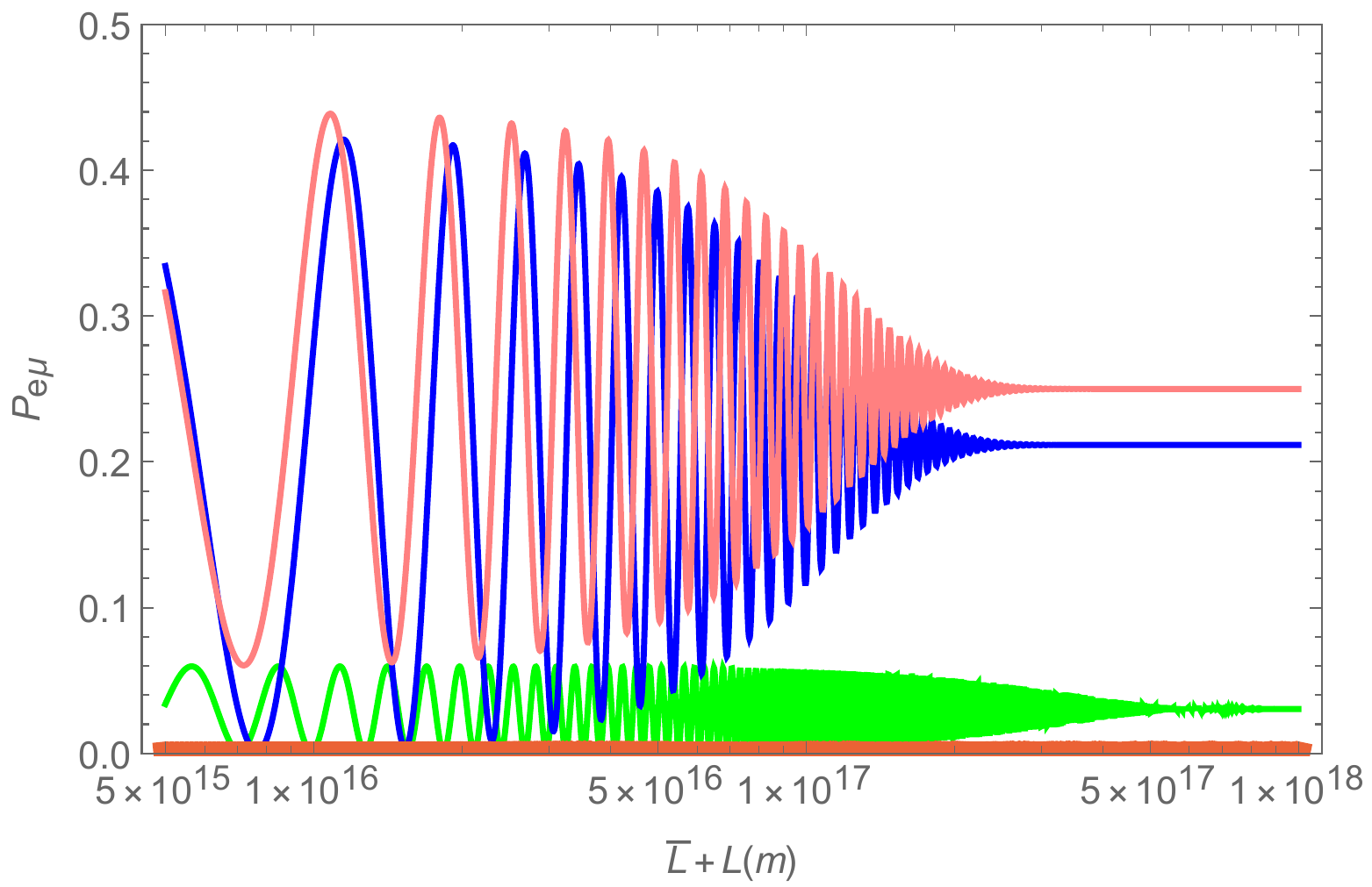}}
	\caption{Probabilities of the survival ({\bf a}) and transition ({\bf b}) patterns between detectors versus their distance $L+\bar L$. The matter potential is taken as $V=0$ (blue), $V=3.208\times 10^{-16} \text{eV}$ (pink), $V=2.087\times 10^{-15}\text{eV}$ (green) and $V=5\times 10^{-14}\text{eV}$ (brown).}\label{fig1}
\end{figure}

For greater clarification, we illustrate the probabilities of the survival (a) and transition patterns (b) between detectors versus their distance $L+\bar L$ in Fig. \ref{fig1}.
Specifically, we plot the probability in terms of $L+\bar L$ for four cases:
vacuum ($V=0$, blue curve), for a value of the potential which corresponds to the neutrino resonance mixing ($V=3.208\times 10^{-16} \text{eV}$, pink curve), for a value of the potential in which ${{\bf L}^m}_{12}^{\text{coh}}$ takes the maximum value ($V=2.087\times 10^{-15}\text{eV}$, green curve) and for a matter-dominated potential ($V=5\times 10^{-14}\text{eV}$, brown curve). 

\section{Quantum Correlations}\label{333} 
As mentioned, the NC neutrino oscillation originates from the neutrino and antineutrino entanglement and quantum coherence due to the overlap of mass eigenstate wave functions. Both of these correlations are affected by the disentangling and decoherence due to the separation of wave packets as well as the scattering of neutrinos and antineutrinos off material medium. Therefore, in this section, we use concurrence and $l_1\text{-norm}$, respectively, as a measure for entanglement and quantum coherence in the framework of quantum resource theory, in order to investigate the mentioned effects.

Entanglement as a quantum correlation is studied extensively in the framework of the quantum resource theory \cite{Vedral,Horodecki,hill,wootters}. Concurrence is one of the measures suggested to quantify entanglement. First, let us define $\tilde \rho$ as follows: 
\be
\tilde{\rho}=(\sigma_y\otimes\sigma_y)\rho^*(\sigma_y\otimes\sigma_y).
\ee
The concurrence measure is generally defined by \cite{hill,wootters}: 
\be
{\cal C}(\rho)=\text{max}(\lambda_1-\lambda_2-\lambda_3-\lambda_4,0),\label{difcon}
\ee
where $\lambda_i$'s are the square roots of the four eigenvalues of the non-Hermitian matrix $\rho\tilde\rho$ in decreasing order.
Accordingly, considering a general two-qubit state, we have
\be
\left|\phi\right\rangle=\alpha_{00} \left|0,0\right\rangle+\alpha_{01} \left|0,1\right\rangle+\alpha_{10} \left|1,0\right\rangle+ \alpha_{11}\left|1,1\right\rangle.\label{gstate}
\ee
The concurrence for this state is obtained as follows
\be
{\cal C}(\rho)=2\mid\alpha_{00}\alpha_{11}-\alpha_{01}\alpha_{10}\mid.
\ee
Clearly, for Bell states, we have ${\cal C}(\rho)=1$. Otherwise, if the coefficients in Eq. (\ref{gstate}) are such that $\left|\phi\right\rangle$ is separable, ${\cal C}(\rho)$ will be zero. For other cases, one obtains a value between 0 and 1. The more entangled the state, the closer the value of ${\cal C}(\rho)$ is to 1.

For an entangled neutrino and antineutrino state originating from the $Z_0$ decay,  the corresponding bipartite state is given in Eq. (\ref{so}) which is written based on the wave packet approach. The density matrix operator corresponding to this state, after integrating over the related momenta and time \footnote{Similar to the calculation of the oscillation probability, we integrate over the propagation time because in all existing neutrino oscillation experiments only the source detector distance is known.}, is obtained as
\bea\nonumber
\rho({\bf x},\bar {\bf x})&\propto&\dfrac{1}{2}\sum_{i,j} \exp\Bigg[-\dfrac{1}{8\sigma_x^2(({\bf v_i}+\bar {\bf v_i})^2+({\bf v_j}+\bar {\bf v_j})^2)}({\bf x}+\bar{\bf x})^2(\Delta {\bf v}_{ij}+\Delta \bar {\bf v}_{ij})^2\\\nonumber
&+&\dfrac{i}{2}(\Delta {\bf p}_{ij}+\Delta  \bar {\bf p}_{ij})({\bf x}+\bar{\bf x})-\dfrac{\sigma_x^2}{2}((\bar {\bf p_i}-{\bf p_i})^2+(\bar {\bf p_j}-{\bf p_j})^2)\\\nonumber
&-&\dfrac{1}{8\sigma_x^2(({\bf v_i}+\bar {\bf v_i})^2+({\bf v_j}+\bar {\bf v_j})^2)}[({\bf x}+\bar {\bf x})({\bf v_i}+{\bf v_j}+\bar {\bf v_i}+\bar {\bf v_j})]^2\\\nonumber
 &+&\dfrac{1}{8\sigma_x^2(({\bf v_i}+\bar {\bf v_i})^2+({\bf v_j}+\bar {\bf v_j})^2)}[-4i\sigma_x^2(\Delta E_{ij}+\Delta\bar E_{ij})+({\bf x}+\bar {\bf x})({\bf v_i}+{\bf v_j}+\bar {\bf v_i}+\bar {\bf v_j})\\
 &+& 2i\sigma_x^2(\bar {\bf p_j}-{\bf p_j})(\bar {\bf v}_j-{\bf v}_j)-2i\sigma_x^2(\bar {\bf p_i}-{\bf p_i})(\bar{\bf v}_i- {\bf v}_i)]^2\Bigg]\mid \nu_i,\bar\nu_i\rangle\langle \nu_j,\bar\nu_j\mid.
\eea
Taking into account the issues related to energies, momenta and group velocities, given in Eqs. (\ref{em})-(\ref{vmbar}), and using the definitions given in Eqs. (\ref{osc}) and (\ref{coh}), one can simplify this relation as follows:
\bea\nonumber
\rho({\bf x},\bar {\bf x})\propto \dfrac{1}{2}\sum_{i,j}&\!\!\!\exp\!\!\!&\Bigg[-2\pi i\dfrac{({\bf x}+\bar{\bf x})}{{{\bf L}^m}_{ij}^{\text{osc}}}-\dfrac{1}{2}\Bigg(\dfrac{{\bf x}+\bar{\bf x}}{{{\bf L}^m}_{ij}^{\text{coh}}}\Bigg)^2-\Bigg(\dfrac{2\pi \xi \sigma_x}{{{\bf L}^m}_{ij}^{\text{osc}}}\Bigg)^2\\
&-\!\!\!\!\!&\dfrac{\sigma_x^2}{2}(\xi-1)^2[(\delta E_i^m- \delta\bar E_i^m)^2+(\delta E_j^m-\delta \bar E_j^m)^2]\Bigg]\mid \nu_i,\bar\nu_i\rangle\langle \nu_j,\bar\nu_j\mid,\label{density1}
\eea
The behavior of this density matrix in terms of ${\bf x}+\bar {\bf x}$ is analogous to that obtained in Ref. \cite{guintidensity} with some differences coming from the distinct propagation of neutrinos and antineutrinos in a material medium.  

\begin{figure}[ht]
 	\centering
 	{\includegraphics[scale=0.5]{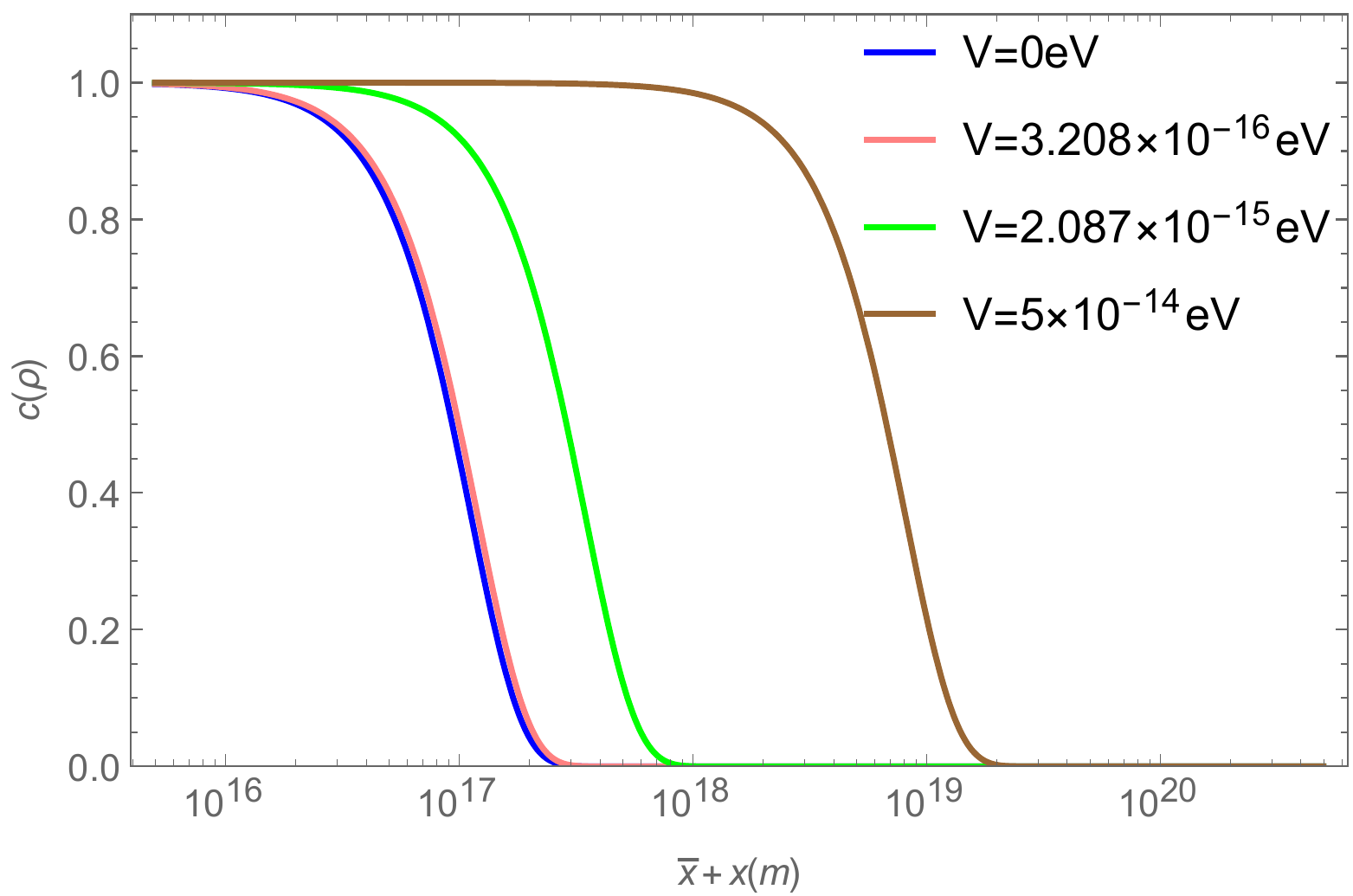}}
 	\caption{Concurrence versus propagation length for various matter potentials. The second and third values of potential have been chosen to correspond to the resonance and the infinite coherence length of a neutrino, respectively.}\label{fig2}
 \end{figure}

Now we can calculate the concurrence whose definition is given in Eq. (\ref{difcon}) as a measure of the neutrino and antineutrino entanglement. By direct calculation, one can obtain the following expression for concurrence:
\be
{\cal C}(\rho)=\exp\Bigg[-\dfrac{1}{2}\Bigg(\dfrac{{\bf x}+\bar{\bf x}}{{{\bf L}^m}_{12}^{\text{coh}}}\Bigg)^2-\Bigg(\dfrac{2\pi \xi \sigma_x}{{{\bf L}^m}_{12}^{\text{osc}}}\Bigg)^2-\dfrac{\sigma_x^2}{2}(\xi-1)^2[(\delta E_1^m- \delta\bar E_1^m)^2+(\delta E_2^m-\delta \bar E_2^m)^2]\Bigg].\label{concurrence}
\ee
In fact, the first term in the exponential has a decisive role in the behavior of ${\cal C}(\rho)$ when the propagation length is of the order of the coherence length.  
Furthermore, we should emphasis that if one uses a plane wave description for neutrinos, the value of concurrence remains equal to 1 for every value of ${\bf x}+\bar{\bf x}$ and $V$. 
 Meanwhile, in the wave packet approach, the concurrence depends on $V$ as well as ${\bf x}+\bar {\bf x}$; it becomes zero when ${\bf x}+\bar {\bf x}$ exceeds the coherence length, and its falling is postponed for higher matter density (see Fig.  \ref{fig2}). 

\begin{figure}[ht]
	\centering
	{\includegraphics[scale=0.5]{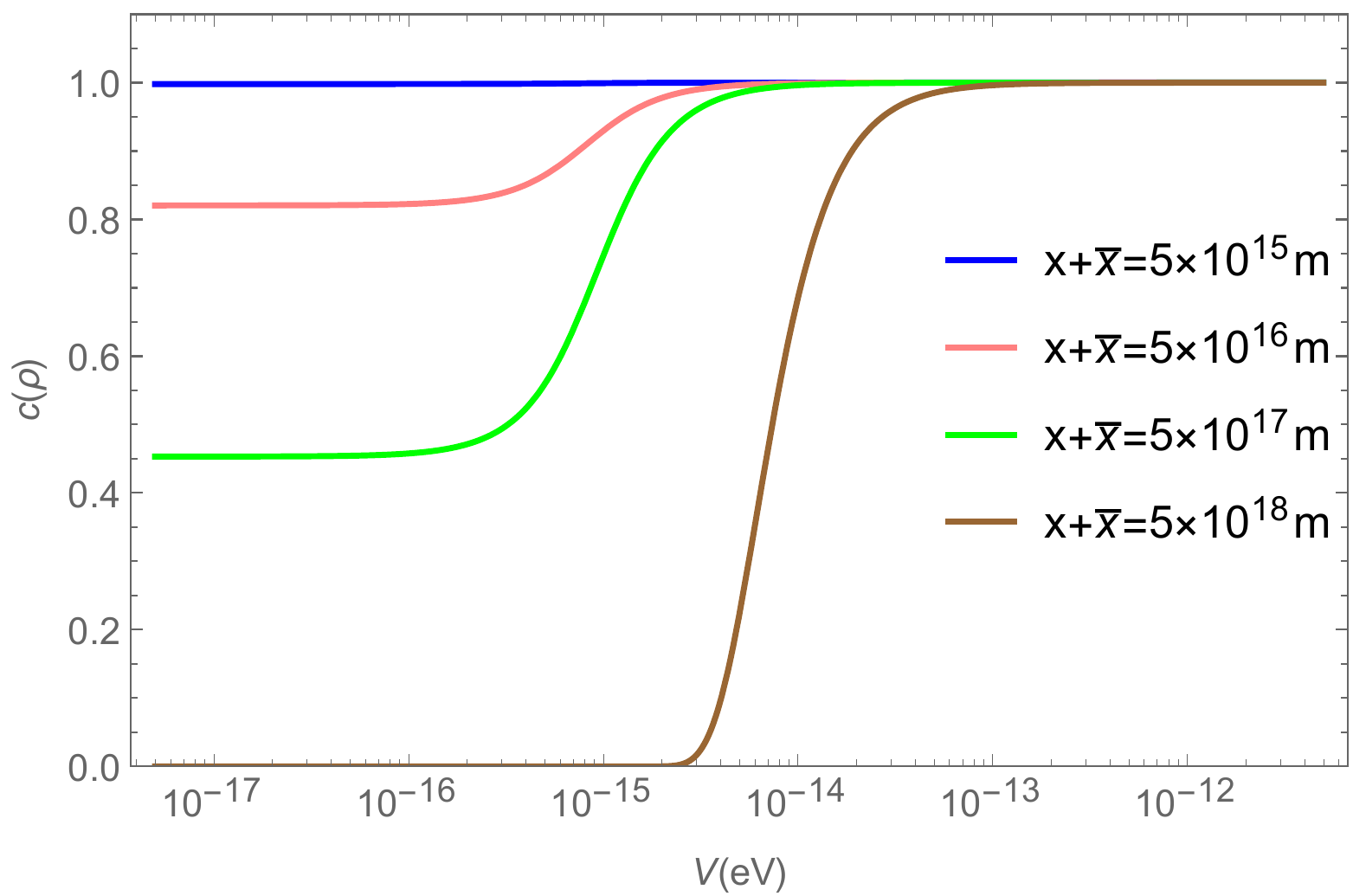}}
	\caption{Concurrence versus  matter potential for various propagation lengths.}\label{fig3}
\end{figure}

 Specifically, according to Eq. (\ref{concurrence}), when the localization properties are considered by the wave packet approach, the overlap of mass eigenstate wave packets is diminished for a propagation length larger than the coherence length because the corresponding velocities are different. However, it can be compensated by propagation of the neutrino in the material medium. 
 In other words, the refraction of neutrinos in a dense medium causes the difference in wave packet velocities to decrease.  While the concurrence vanishes for smaller matter potentials for a specific propagation length, it becomes nonzero for larger potentials; for example, see Fig. \ref{fig3}.

Now let us study the coherence of neutrinos and antineutrinos originating from the $Z_0$ decay process. As was noted, in order for oscillation to occur, neutrinos must be coherent during the production, propagation, and detection processes. 
In the framework of the quantum resource theory, quantum coherence is a quantum correlations that can be quantified by appropriate measure quantities \cite{bau,str}.
Among the several measures for quantum coherence, a very
intuitive one is related to the off-diagonal elements of the considered quantum state. Hence, for a given density matrix, $\rho$, the $l_1\text{-norm}$ as a measure of quantum coherence is defined by 
\be\label{l1}
\textit{c}(\rho)=\sum_{i\neq j}|\rho_{ij}|.
\ee 
In general, the maximum possible value for $ \textit{c}(\rho)$ is  $d-1$, where $d$ is the dimension of the corresponding density matrix \cite{str}. Therefore, the neutrino oscillation shows that the off-diagonal elements of the corresponding density matrix in the basis of the flavor state are nonzero, and consequently we have no vanishing $l_1-\text{norm}$ \cite{epjp}.

\begin{figure}[ht]
	\centering
	{\includegraphics[scale=0.5]{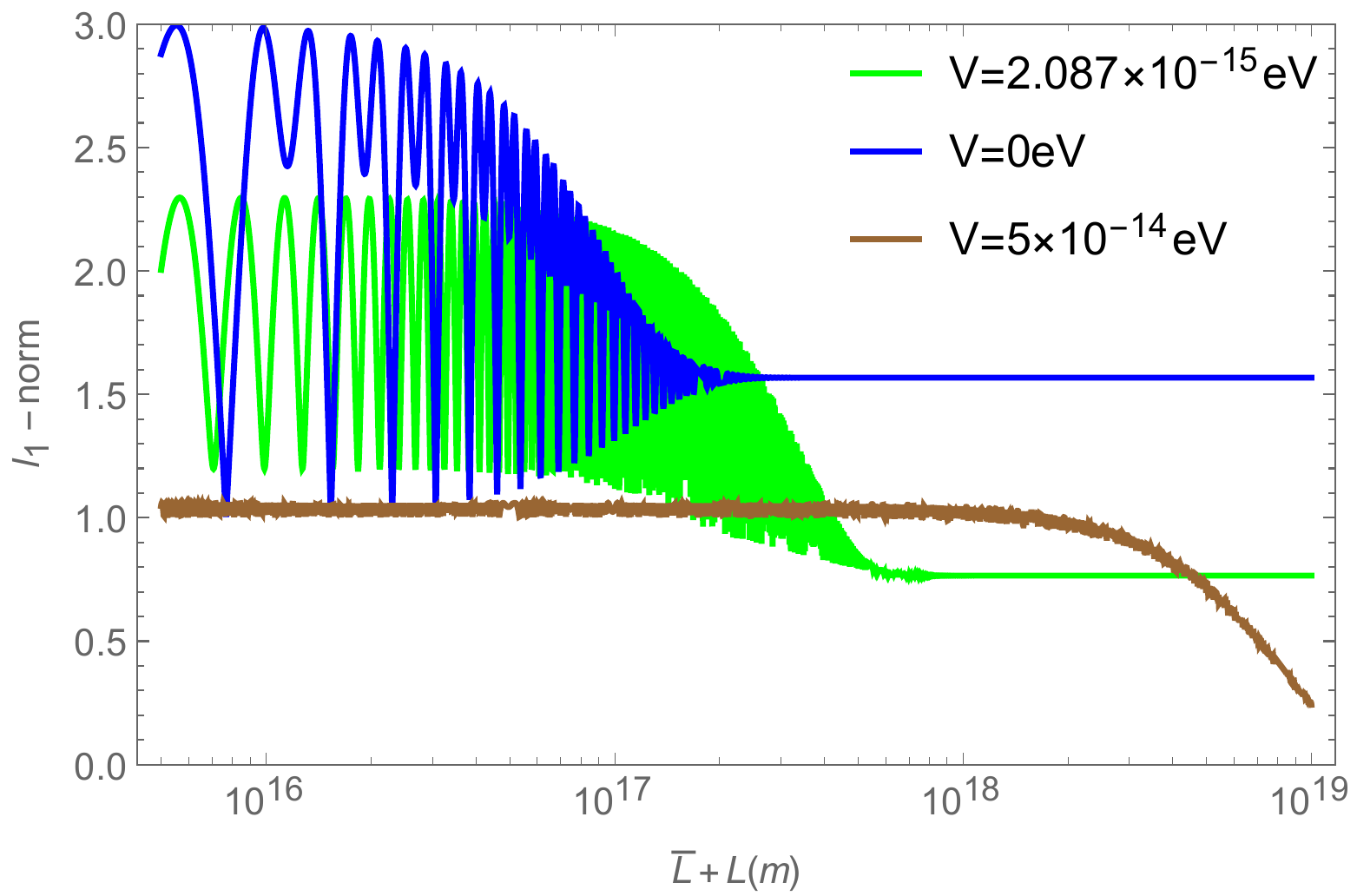}}
	\caption{$l_1$-norm versus the baseline $\bar x+x$ for three values of matter potential.}\label{fig4}
\end{figure}

In the case of NC neutrinos, the dimension of the density matrix given in Eq. (\ref{density1}) is four in two-flavor schema. In this case, therefore, the maximum value of the $l_1$-norm is 3. According to the definition given in Eq. (\ref{l1}),  
one can write this parameter in terms of the transition amplitudes as follows:
\bea\nonumber
\textit{c}(\rho)&=&2\Bigg\{|A_{ee}^m({\bf x},\bar {\bf x})A_{e\mu}^{m*}({\bf x},\bar {\bf x})| +|A_{ee}^m({\bf x},\bar {\bf x})A_{\mu e}^{m*}({\bf x},\bar {\bf x})| +|A_{ee}^m({\bf x},\bar {\bf x})A_{\mu\mu}^{m*}({\bf x},\bar {\bf x})| \\
&+&|A_{e\mu}^m({\bf x},\bar {\bf x})A_{\mu e}^{m*}({\bf x},\bar {\bf x})| +|A_{e\mu}^m({\bf x},\bar {\bf x})A_{\mu \mu}^{m*}({\bf x},\bar {\bf x})|+|A_{\mu e}^m({\bf x},\bar {\bf x})A_{\mu \mu}^{m*}({\bf x},\bar {\bf x})| 
  \Bigg\}.
\eea
Here, the first and second terms and the fifth and sixth terms are identical. 
An explicit form of $\textit{c}(\rho^m)$ in terms of neutrino oscillation parameters is given in the Appendix. We illustrate the behavior of $\textit{c}(\rho^m)$ versus the distance of detectors ${\bf x}+\bar {\bf x}$ through Fig. \ref{fig4}. 
In this plot, we consider three values for matter potential: vacuum ($V=0$),  
 the value corresponding to the infinite coherence length for neutrinos, and a matter-dominated value. 
 We see that although for baseline less than the coherence length the matter effect causes $\textit{c}(\rho^m)$ to drop relative to vacuum, the coherence length in a material medium becomes larger than one in the vacuum. 

\begin{figure}[ht]
	\centering
	\subfigure[]{\includegraphics[scale=0.5]{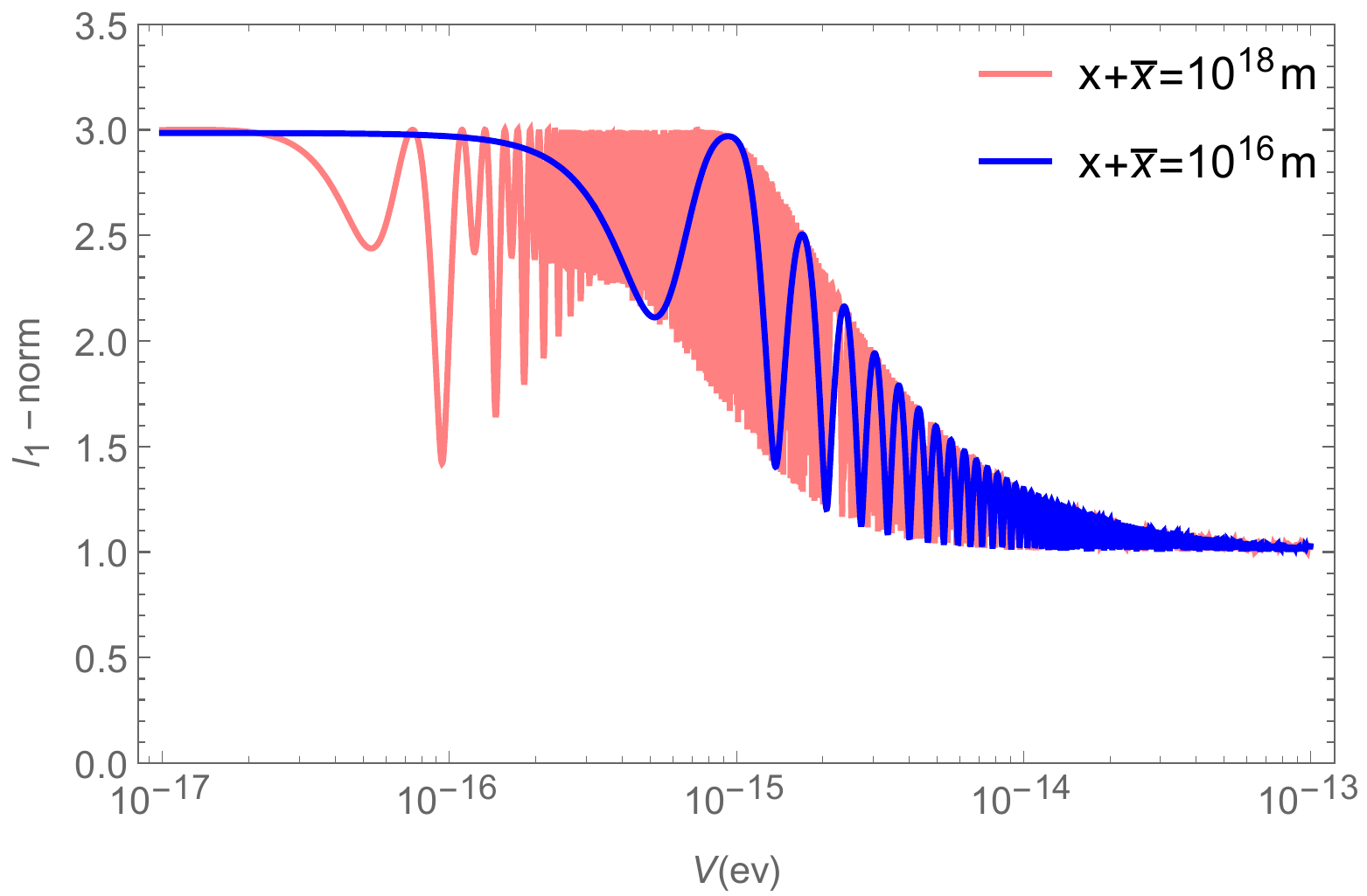}}\subfigure[]{\includegraphics[scale=0.5]{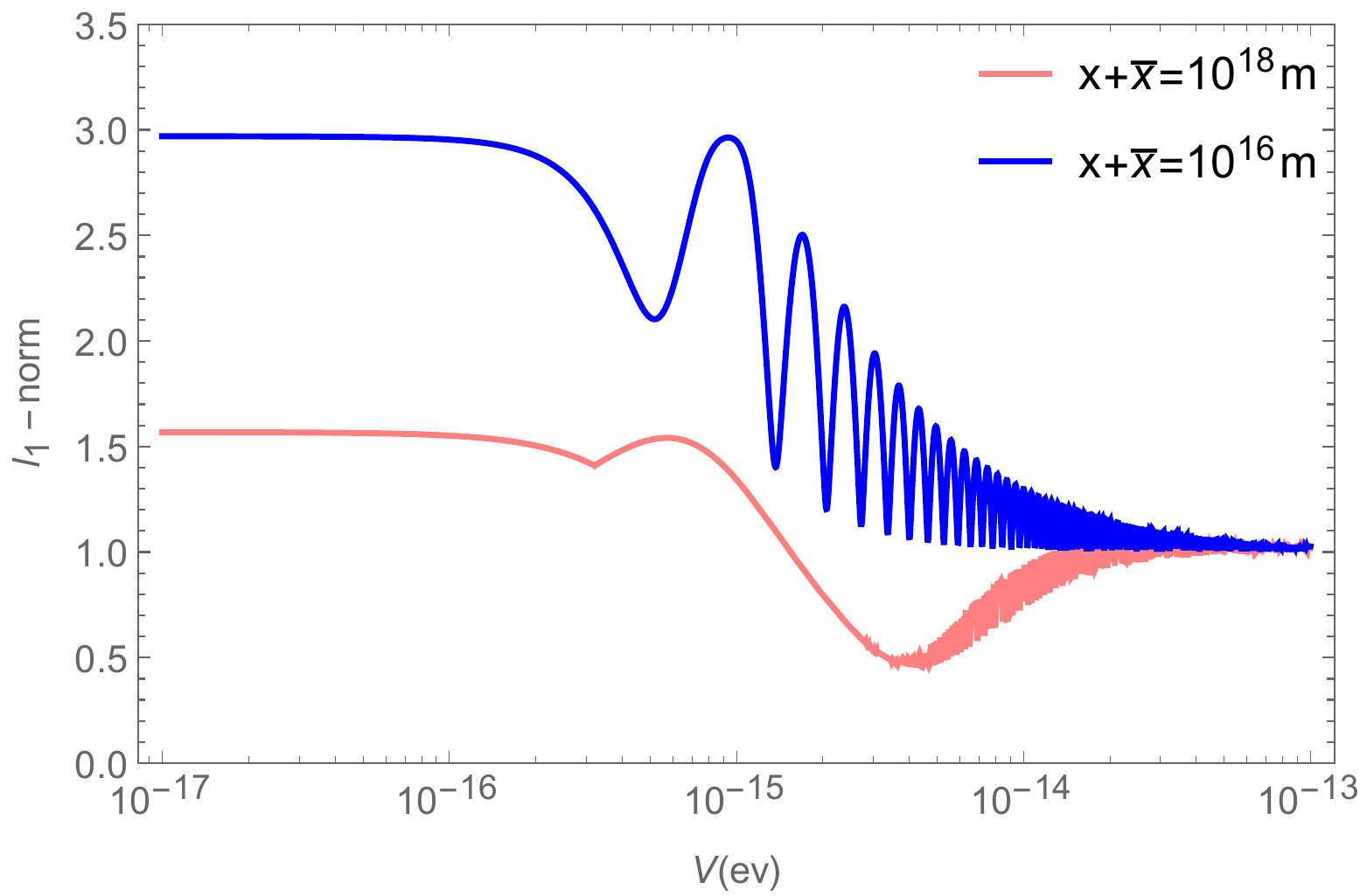}}
	\caption{$l_1$-norm versus the matter potential in both the plane wave (a) and wave packet approaches (b). Two values are taken for the baseline; a value smaller and a value larger than the coherence length.}\label{fig5}
\end{figure}

Furthermore, in Fig. \ref{fig5}, the behavior of $\textit{c}(\rho^m)$ in terms of the matter potential is depicted  for the results obtained based on both plane wave (a) and wave packet (b) approaches. We consider two values for the baseline: a value smaller than the coherence length, $10^{16}$m (blue curve), and a value greater than that, $10^{18}$m (pink curve). In the case of the former (blue curve), $\textit{c}(\rho^m)$ obtained in
both approaches behaves similarly. Although for some values of the potential a slight increase in the value of $\textit{c}(\rho^m)$ is seen, as the value of the potential increases,  $\textit{c}(\rho^m)$ eventually decreases to a value less than the corresponding one in the vacuum. However, in the case of $10^{18}$m baseline (pink curve), we see that in the plane wave approach, $\textit{c}(\rho^m)$ achieves the maximum value for a region of the potential value, while in the wave packet approach, $\textit{c}(\rho^m)$ is certainly smaller than the maximum value. This is due to the decoherence effects coming from the separation of neutrino wave packets.

\section{Conclusion}
According to the theory of neutrino oscillation, it is possible to observe an oscillation pattern between the detectors of neutrinos and antineutrinos coming from $Z_0$ boson decay  when the neutrino production, propagation, and detection coherence conditions are satisfied. We see that under a realistic condition for $Z_0$ boson decay, $\sigma_x$ is estimated to be about $10^{-16}$m. Thus the coherence length is smaller than the oscillation length, and the propagation coherence conditions are not satisfied. Nevertheless, we have assumed $\sigma_x=5\times10^{-10}$ so that the coherence conditions are satisfied.
However, if only one of them is detected, neutrino oscillation cannot be observed. This phenomenon results from the fact that the neutrino and antineutrino arising from the $Z_0$ decay are both entangled and coherent. These quantum properties can be quantified according to the quantum resource theory. In this paper, we use the concurrence and $l_1$-norm to quantify entanglement and quantum coherence, respectively. On the other hand, considering localization properties by using the wave packet approach, one can see that the wave packet separation of neutrino mass eigenstates due to the difference in the corresponding velocities suppresses the amount of entanglement and coherence measures. Another factor that can affect these quantum correlations is the propagation of neutrinos and antineutrinos in material media. Hence, in this paper, we have reanalyzed the  $Z_0$ decay neutrino oscillation by considering the localization properties and matter potential. In particular, we have obtained the concurrence and $l_1$-norm in two-flavor schema. If the localization effects are ignored and in other words, the plane wave approach is used to obtain the transition probability, the concurrence always remains 1. Otherwise, the concurrence ceases when the propagation length exceeds the coherence length. However, the matter potential causes the coherence length to increase, and as a result, this damping occurs in larger propagation lengths (see Fig. \ref{fig2}). Indeed, the propagation in material medium causes wave packet separation, with the result that the damping of entanglement is compensated to some extent (see Fig. \ref{fig3}). Furthermore, we have shown that although the matter potential causes the $l_1$-norm value to be small compared to the corresponding one in the vacuum, the damping due to the wave packet separation occurs in larger propagation lengths as well (see Fig. \ref{fig4}).  As another point, we have compared the behavior of the $l_1$-norm in terms of matter potential for the case when using the plane wave approach and when using the wave packet approach. Obviously, when the propagation length is smaller than the coherence length, both approaches give the same result. Otherwise, in the plane wave approach, the coherence value is maintained, but it is ultimately suppressed due to the interaction with the material medium. In the wave packet approach, its value is suppressed before the significant impact of the potential (see Fig. \ref{fig5}).

\appendix
\section{}
Using Eq. (\ref{density1}) for density matrix, one can obtain the following expression for the $l_1$-norm in the limit of a large baseline: 
\bea\nonumber
c(\rho)&=&\sum_{i=1}^{2}\sqrt{h^2_{12}\Omega_i^2-\dfrac{1}{4}\gamma_1^2(h_{12}^2\sin^2(2\pi f_{12})-1)-\gamma_1\Omega_i h_{12}\cos(2\pi f_{12})}\\\nonumber
&+&\sqrt{\gamma_2^2+h_{12}^2\sin^22\theta^m-\gamma_1^2h_{12}^2\cos^2(2\pi f_{12})-2\gamma_1\gamma_4h_{12}\cos(2\pi f_{12})}\\
&+&\sqrt{\gamma_3^2+h_{12}^2\sin^22\bar\theta^m-\gamma_1^2h_{12}^2\cos^2(2\pi f_{12})-2\gamma_1\gamma_4h_{12}\cos(2\pi f_{12})},
\eea
in which
\be
h_{12}=\exp[-\dfrac{1}{2}(\dfrac{{\bf x}+\bar{\bf x}}{{{\bf L}^m}_{12}^{\text{coh}}})^2-(\dfrac{2\pi \xi \sigma_x}{{{\bf L}^m}_{12}^{\text{osc}}})^2]
\ee
\be
f_{12}=\dfrac{{\bf x}+\bar{\bf x}}{{{\bf L}^m}_{12}^{\text{osc}}}
\ee
and
\be
\Omega_1
=\cos^2\theta^m\sin^2\bar\theta^m+\sin^2\theta^m\cos^2\bar\theta^m,
\ee
\be
\Omega_2
=-\left(\cos^2\theta^m\cos^2\bar\theta^m+\sin^2\theta^m\sin^2\bar\theta^m\right),
\ee

\be
\gamma_1=\sin2\theta^m\sin2\bar\theta^m,
\ee
\be
\gamma_2=\cos2\theta^m\sin2\bar\theta^m,
\ee
\be
\gamma_3=\sin2\theta^m\cos2\bar\theta^m,
\ee
\be
\gamma_4=\cos2\theta^m\cos2\bar\theta^m.
\ee


\begin{thebibliography}{32}
\bibitem{LG2016}
J.A. Formaggio, D.I. Kaiser, M.M. Murskyj, T.E. Weiss, Phys. Rev. Lett. {\bf 117}, 050402 (2016).
\bibitem{c1}
X.-K. Song, Y. Huang, J. Ling, M.-H.
Yung, Phys. Rev. A {\bf 98}, 050302(R) (2018).
\bibitem{b1}
M. Blasone, F. Dell’Anno, S. De Siena,
F. Illuminati, Europhys. Lett. {\bf 85}, 50002 (2009).
\bibitem{b2}
M. Blasone, F. Dell’Anno, S. De Siena,
F. Illuminati, Europhys. Lett. {\bf 112}, 20007 (2015).
\bibitem{epl}
M.M. Ettefaghi, Z.S. Tabatabaei Lotfi, R. Ramezani
Arani, Europhys. Lett. {\bf 132}, 31002 (2020).
\bibitem{e2}
A.K. Alok, S. Banerjee, S. U. Sankar, 
Nucl. Phys. B {\bf 909}, 65 (2016).
\bibitem{e3}
S. Banerjee, A.K. Alok, R. Srikanth,
B.C. Hiesmayr, Eur. Phys. J. C {\bf 75}, 1 (2015).
\bibitem{e4}
J. Naikoo, A.K. Alok, S. Banerjee,
S.U. Sankar, G. Guarnieri, C. Schultze,
B.C. Hiesmayr, Nucl.
Phys. B {\bf 951}, 114872 (2020).
\bibitem{epjp}
Z. Askaripour Ravari, M.M. Ettefaghi, S. Miraboutalebi, Eur. Phys. J. Plus {\bf 137}, 488 (2022).
\bibitem{prd}
M.M. Ettefaghi, R. Ramazani Arani, Z.S. Tabatabaei Lotfi, Phys. Rev. D {\bf 105}, 095024 (2022).
\bibitem{11}
Y.W. Li, L.J. Li, X.K. Song, D. Wang, L.Ye,  Eur. Phys. J. C {\bf 82}, 799(2022). 
\bibitem{12}
A.K. Jha, A. Chatla, Eur. Phys. J. Spec. Top. {\bf 231}, 141 (2022).
\bibitem{13}
B. Yadav, T. Sarkar, K. Dixit, A.K. Alok, Eur. Phys. J. C {\bf 82}, 1 (2022).
\bibitem{43}
B. Kayser, Phys. Rev. D {\bf 24}, 110 (1981).
\bibitem{44}
M. Beuthe, Phys. Rep. {\bf 375}, 105 (2003).
\bibitem{45}
E.Kh. Akhmedov, A.Yu. Smirnov, Phys. At. Nucl. {\bf 72}, 1363 (2009). 
\bibitem{46}
C. Giunti, C.W. Kim, Phys. Rev. D {\bf 58}, 017301 (1998).
\bibitem{found}
E.Kh. Akhmedov, A.Yu. Smirnov, Found. Phys. {\bf 41}, 1279 (2011).
\bibitem{glashow}
A.G. Cohen, S.L. Glashow, Z. Ligeti, Phys. Lett. B {\bf 678}, 191 (2009).
\bibitem{a2012}
 E.Kh. Akhmedov, D.Hernandez, A.Yu Smirnov, J. High Energy Phys. {\bf 04}, 1 (2012).
\bibitem{guintidensity}
C. Giunti, Found. Phys. Lett. {\bf 17}, 103 (2004).
\bibitem{sza}
A.Yu. Smirnov, G.T. Zatsepin,  Mod. Phys. Lett. A{\bf 7} 1272 (1992).
\bibitem{4}
M.M. Ettefaghi, Z. Askaripour Ravari, Phys. Lett. B {\bf 59}  747 (2015).
\bibitem{scrip}
M.M. Ettefaghi, Z. Askari Ravari. Phys. Scr. {\bf 95} 035301 (2020).
\bibitem{dds}
M. Kaur, M. Singh, Sci. Rep. {\bf 10}, 11427 (2020).
\bibitem{wolf}
L. Wolfenstein, Phys. Rev. D {\bf 17}, 2369 (1978). 
\bibitem{mikh}
S.P. Mikheyev, A.Yu. Smirnov, Sov. J. Nucl. Phys. {\bf 42}, 913 (1985).
\bibitem{1}
P.B. Denton, H. Minakata, S.J. Parke, J. High Energy Phys. {\bf 1606}, 051 (2016).
\bibitem{2000qm}
C. Giunti, C.W. Kim, Found. Phys. Lett. {\bf 14}, 213 (2001).
\bibitem{giu02}
C. Giunti, J. High Energy Phys. {\bf 0211}, 017 (2002).
\bibitem{Vedral}
V. Vedral, M.B. Plenio, M.A. Rippin,  P.L. Knight, Phys. Rev. Lett. {\bf 78}, 2275 (1997).
\bibitem{Horodecki}
R. Horodecki, P. Horodecki, M. Horodecki, K. Horodecki, Rev. Mod. Phys. {\bf 81}, 865 (2009).
\bibitem{hill}
S.A. Hill, W.K. Wootters, Phys. Rev. Lett. {\bf 78},  5022 (1997).
\bibitem{wootters}
W.K. Wootters, Phys. Rev. Lett. {\bf 80},  2245 (1998).
 \bibitem{bau}
T. Baumgratz,  M. Cramer, M.B. Plenio.  Phys. Rev. Lett. {\bf 113}, 140401 (2014).
\bibitem{str}
A. Streltsov, G. Adesso, M.B. Plenio. Rev. Mod. Phys. {\bf 89}, 041003 (2017).
\bibitem{2020e}
K. Abe et al., Phys. Rev. Lett. {\bf 124}, 161802 (2020).

	


	


	  
	 


\end{thebibliography}
\end{document}